\documentclass[11pt]{article}

\usepackage[margin=1in]{geometry}
\usepackage{times}
\usepackage{amsmath,amssymb}
\usepackage{graphicx}
\usepackage{xurl}   
\usepackage{hyperref}
\usepackage{booktabs}
\usepackage{enumitem}
\usepackage[numbers]{natbib}
\usepackage{xcolor}

\title{\bfseries
Can Dynamic Spectrum Sharing Protect Passive Radio Sciences?
}

\author{
Gregory Hellbourg\\
California Institute of Technology
}

\date{}

\begin{document}
\maketitle

\begin{abstract}
Dynamic Spectrum Sharing (DSS) is increasingly promoted as a central element of modern spectrum policy, driven by growing demand from commercial wireless systems, perceptions of spectrum underutilization, and advances in sensing and spectrum access technologies. Passive radio sciences, including radio astronomy, Earth remote sensing, and meteorology, operate under fundamentally different constraints. They depend on access to exceptionally low interference spectrum and are uniquely vulnerable to even brief or intermittent radio frequency interference (RFI).

In this white paper, we examine whether DSS can offer meaningful benefits to passive services, or whether it primarily introduces new failure modes and enforcement challenges that undermine reliable protection. We introduce the concept of ``just-in-time quiet zones'' (JITQZ) as a targeted mechanism for protecting high value observations within dynamic spectrum environments, and we assess hybrid regulatory frameworks that preserve static protection for core passive bands while allowing carefully constrained dynamic access in adjacent frequency ranges. We analyze the limiting role of radio frequency propagation uncertainty, electromagnetic compatibility (EMC) constraints, and the current lack of comprehensive spectrum situational awareness.

To formalize these challenges, we develop a game theoretic framework for DSS for passive services. We show why non cooperative sharing fails in ``one shot'' settings, identify the conditions under which cooperation can be sustained in repeated interactions with imperfect monitoring, and examine incentive mechanisms such as spectrum credits, penalties, and pseudonymetry-enabled attribution that can shift behavior toward compliance. 

We conclude that DSS can support passive radio sciences only when treated as a high reliability, safety critical system. Static spectrum allocations and permanent quiet zones remain indispensable for the most sensitive observations. Dynamic access may be considered in extended bands only when supported by conservative propagation aware safeguards, robust EMC practices, credible attribution, and enforceable accountability.
\end{abstract}

\tableofcontents
\newpage

\section{Introduction}

\subsection{Dynamic Spectrum Sharing}

Dynamic Spectrum Sharing (DSS) refers to regulatory and technical frameworks in which access to radiofrequency spectrum resources are allocated dynamically rather than through long term, exclusive assignments. In a DSS regime, authorization to transmit may vary over time, frequency, geographic location, antenna pointing, or transmit power, depending on operational context and policy constraints. This represents a significant departure from traditional spectrum management, which has historically relied on fixed allocations and service specific frequency bands that remain in place for decades.

Modern implementations of DSS are typically enabled by a combination of software defined radios, spectrum access databases, sensing capabilities, and automated enforcement mechanisms. These systems are designed to support coexistence among multiple categories of users, including commercial operators, federal systems, scientific services, and safety-of-life applications, each operating under different priority and protection rules. In this sense, DSS is not merely a technical innovation but a true structural shift in how spectrum rights and responsibilities are defined and exercised.

DSS is often justified as a response to the perceived imbalance between rapidly increasing demand for wireless services and the uneven utilization of spectrum across space and time. In dense metropolitan areas, where congestion is real and measurable, the motivation is straightforward. Expanding access to spectrum is closely associated with economic growth, increased competition, and improved connectivity. However, spectrum occupancy is highly heterogeneous. In many rural or remote regions, substantial portions of the spectrum remain naturally quiet under existing static allocations. This geographic asymmetry is a critical consideration for passive radio sciences, which depend on these quiet environments to achieve their scientific objectives.

\subsection{Policy effort in the United States}

The policy momentum behind spectrum sharing is explicit and sustained. In the United States, spectrum sharing has been promoted at the highest levels of government for more than a decade. A 2018 presidential memorandum on developing a sustainable spectrum strategy directed federal agencies to pursue more efficient spectrum use and explicitly encouraged greater sharing between federal and non-federal users \cite{wh2018_sustainable_spectrum}. In 2023, a subsequent presidential memorandum initiated the development of a National Spectrum Strategy \cite{fr2023_memo}, and the final Strategy document emphasized dynamic spectrum sharing as a key mechanism for meeting growing demand while maintaining support for federal missions \cite{ntia2023_nss_pdf}. This direction was further reinforced in a 2025 National Security Presidential Memorandum on next-generation wireless leadership, which highlighted the need for flexible spectrum access frameworks to support future systems \cite{wh2025_6g_memo}.

Although the specific language and emphasis vary across these documents, the broader trajectory is clear. U.S. spectrum policy is moving away from static, service specific allocations and toward more dynamic and technological approaches. This shift raises fundamental questions about how services with vastly different tolerance to interference can coexist within such frameworks.

\subsection{Passive radio sciences}

Passive radio sciences differ from active spectrum users in a fundamental way. They do not transmit signals into the environment but instead measure naturally occurring electromagnetic emissions or reflections originating from astronomical objects or geophysical processes. These signals are often extraordinarily faint, frequently many orders of magnitude weaker than the emissions produced by modern communication systems \cite{Hellbourg2025QuietSkies}.

Because passive systems do not control the strength of the signals they observe, they cannot compensate for interference by increasing transmit power or changing modulation schemes. Sensitivity is achieved through long integration times, large collecting areas, wide instantaneous bandwidths, high spectral resolution, and extremely low-noise receiver chains. Modern instruments are designed to operate near fundamental physical limits in order to push the boundaries of what can be observed in the Universe or measured in the Earth system.

This extreme sensitivity introduces two distinct vulnerabilities. First, even brief or intermittent interference can corrupt scientific data in irreversible ways. Interference can bias calibration solutions, reduce effective sensitivity through data excision, or mask transient phenomena that occur on millisecond timescales and cannot be recovered after the fact. Second, strong interference can drive receivers into nonlinear operating regimes. When front-end components saturate, or when analog-to-digital converters exceed their dynamic range, the resulting distortion can affect wide portions of the observed spectrum and persist beyond the duration of the interfering signal itself. In such cases, interference does not merely add noise but fundamentally alters the measurement process.

As radio astronomy and related fields continue to pursue higher sensitivity in order to probe fainter signals and more distant phenomena, tolerance to interference is decreasing. This trend goes against the implicit assumption in many DSS frameworks that systems can gracefully adapt to more dynamic and crowded spectral environments.

The implication is therefore severe. The central question addressed in this paper is not whether DSS can improve average spectrum utilization, but whether it can deliver predictably enforceable quiet conditions that meet the reliability requirements of passive radio sciences.

\subsection{Thesis}

This paper examines dynamic spectrum sharing through the lens of three classes of passive services. Radio astronomy serves as a primary example, encompassing both deep integrations and time-critical transient observations. Passive Earth remote sensing is considered for its role in climate monitoring, hydrology, and oceanography, where protected spectral bands underpin long term environmental records. Meteorology is included due to its reliance on passive microwave observations that feed directly into numerical weather prediction and operational forecasting.

Our central thesis is deliberately cautious. Static spectrum allocations remain essential for the most sensitive passive observations and for long term scientific continuity. Dynamic spectrum sharing may nevertheless offer value in adjacent or extended bands, provided that several stringent conditions are met. Enforcement mechanisms must be credible and reliable, radio frequency propagation uncertainty must be explicitly accounted for rather than idealized, and attribution and accountability must be strengthened. Techniques such as OFDM watermarking for pseudonymetry are highlighted as potential tools to support enforceable coexistence in such dynamic environments.

\section{Value of quiet spectrum for passive radio sciences}

\subsection{Radio astronomy}

Radio astronomy enables observations of neutral hydrogen, pulsars, fast radio bursts, solar and planetary radio emission, cosmic magnetism, and a wide range of fundamental astrophysical processes. Modern radio telescopes are typically not limited by the intrinsic strength of astronomical signals, but by system noise and external radio frequency interference (RFI). As a result, the cost of interference is rarely a simple or transient loss.

Interference can bias calibration solutions in subtle ways, leading to systematic errors that persist even after aggressive data flagging and excision. When contaminated data are removed, sensitivity is reduced in a nonlinear manner for many imaging and synthesis problems, particularly for wide-field instruments and arrays with complex sidelobe structures. In addition, strong or structured interference can imprint persistent artifacts in images, for example through sidelobe contamination that redistributes power across the field of view. For time domain astronomy, the consequences are even more severe because transient events such as fast radio bursts or short lived solar phenomena cannot be replayed or recovered once missed.

Beyond these effects on data products, radio astronomy receivers are themselves vulnerable to strong interference. Modern front-end systems are designed for extreme sensitivity and often operate close to the limits of linearity. Strong RFI can drive low noise amplifiers or downstream components into nonlinear regimes, leading to gain compression, intermodulation products, or saturation of analog-to-digital converters. When this occurs, the impact is not confined to the frequency of the interfering signal. Wide portions of the band can be corrupted simultaneously, and the effects may persist beyond the duration of the interference due to recovery times and calibration instability. This behavior is fundamentally different from the additive noise model often assumed in spectrum sharing discussions.

These vulnerabilities are becoming more pronounced as radio astronomy pushes toward higher sensitivity. The scientific drive to probe fainter signals and more distant epochs of the Universe implies lower noise figures, wider bandwidths, and higher dynamic range requirements. In practical terms, this means that tolerance to RFI is decreasing as the spectral environment becomes more crowded.

\subsection{Earth remote sensing and meteorology}

Passive Earth remote sensing and meteorology rely on precise measurements of naturally emitted microwave and radio frequency radiation to retrieve environmental parameters such as soil moisture, sea surface temperature, atmospheric water vapor, and cloud properties. These measurements are anchored in protected spectral windows and molecular absorption lines that have been carefully selected to maximize sensitivity to specific geophysical variables.

Interference in these bands can introduce retrieval biases, data gaps, or systematic errors that propagate into climate records and numerical weather prediction models. Unlike many commercial systems, these errors may not be immediately apparent, yet they can accumulate over long time periods and degrade the integrity of essential environmental datasets.

In recent years, the expansion of broadband wireless services, including 5G deployments in bands adjacent to passive allocations, has highlighted the fragility of this balance. Even when transmissions nominally comply with allocation boundaries, out-of-band emissions, and aggregate interference have been shown to impact passive sensors, particularly spaceborne instruments with wide fields of view and limited filtering options. These concerns underscore that protection of passive services depends not only on nominal frequency assignments, but also on emission masks, aggregation effects, and real-world deployment practices \cite{English2025RFI,Yousefvand2020Impact5G}.

The value of clean spectrum for remote sensing and meteorology is therefore not abstract or academic. It underpins food security, disaster preparedness, climate monitoring, and water resource management. Any spectrum sharing framework that affects these services must be evaluated against these societal consequences.

\subsection{Underutilized spectrum}

A recurring argument in favor of dynamic spectrum sharing is that large portions of the spectrum appear to be underutilized. From the perspective of passive radio sciences, this argument requires careful qualification. Spectrum occupancy is highly dependent on geography, time, measurement methodology, and propagation conditions. What appears unused in one location or at one time may be heavily occupied elsewhere or under different atmospheric conditions.

Moreover, not all quiet spectrum is interchangeable. Radio astronomy, in particular, relies on a small number of frequency bands that are of exceptional scientific importance because they correspond to fundamental molecular and atomic transitions. The hydrogen 21-cm line is the most famous example, but many other spectral lines are critical tracers of physical conditions in the Universe. These protected bands form the backbone of radio astronomical science and cannot be arbitrarily relocated.

At the same time, radio astronomy routinely observes outside protected bands. Cosmological redshift shifts spectral lines away from their rest frequencies, requiring access to adjacent or entirely different bands in order to probe different epochs of cosmic history. In addition, many observations target broadband continuum emission, where sensitivity improves with increased bandwidth regardless of specific frequencies. The fundamental scaling is captured by the radiometer equation \cite{RohlfsWilson2004}, which shows that sensitivity improves with the square root of bandwidth and integration time. Access to additional quiet spectrum therefore directly translates into improved scientific capability.

This dual reality complicates simplistic notions of spectrum underutilization. Some bands are protected because they are irreplaceable, while others are scientifically valuable precisely because they have historically been avoided due to interference. In practice, frequency ranges used for satellite operations or other active services are often notched out entirely in radio astronomical observations. As a result, entire ranges of redshift space and continuum sensitivity remain unexplored. From a scientific perspective, these gaps correspond to unobserved layers of space and time.

Field measurements further illustrate the importance of baseline conditions. In many remote regions, such as parts of the western United States, the spectrum is already remarkably quiet under existing static allocations. Measurements conducted near urban environments, intermediate rural sites, and remote desert locations reveal orders-of-magnitude differences in RFI occupancy and intensity. In such contexts, the primary objective for passive sciences may not be to enable additional sharing, but to preserve naturally quiet environments that are increasingly rare.

Finally, it is important to note that satellite services occupy a special role in this discussion. Satellite emissions are global by nature, aggregate rapidly, and are often visible even in the most remote locations. Any consideration of DSS for passive sciences must explicitly address whether, and under what conditions, satellite systems are included. Without this clarity, dynamic sharing frameworks risk offering protection only against terrestrial emitters while leaving the dominant interference sources unaddressed.

\section{DSS for passive services}

\subsection{Forms of Dynamic Spectrum Sharing}

Dynamic Spectrum Sharing is not a single, uniform concept but rather a family of approaches that differ in how access decisions are made and enforced. These approaches are often grouped according to the mechanisms used to determine whether transmission is permitted at a given time and location.

In database-driven sharing frameworks, transmitting devices query a centralized spectrum access system that encodes regulatory rules, incumbency protections, and geographic constraints. Access is granted or denied based on predefined policies rather than on real time detection of other users. This approach has the advantage of predictability and centralized control, but it depends critically on the completeness and accuracy of the underlying information.

Sensing-driven sharing relies on devices detecting other users in the band and opportunistically accessing spectrum when no incumbent activity is detected. While appealing in concept, this approach is fundamentally inefficient for protecting passive services. A radio telescope, Earth-observing radiometer, or meteorological sensor does not emit a signal that can be reliably detected by external devices. From the perspective of a sensing-based system, the spectrum may appear unused even when a highly sensitive receiver is actively observing.

Hybrid approaches combine databases with sensing, using centralized rules to establish baseline constraints while allowing local sensing to refine decisions. In principle, this can improve efficiency while maintaining some degree of protection. In practice, the effectiveness of such systems for passive services still depends on conservative rule-based exclusions rather than on detection sensing alone.

Finally, some forms of DSS rely on explicit coordination agreements rather than automated access decisions. These agreements may include time-of-day restrictions, geographic exclusion zones, or operational constraints negotiated between stakeholders. While less flexible than fully dynamic systems, such arrangements can provide clearer protection for sensitive users when properly enforced.

A structural challenge underlies all of these approaches. Spectrum regulation is almost entirely focused on transmitters. Receivers are generally not licensed, registered, or even visible to regulators, and their operating parameters, including sensitivity and dynamic range, are not known to other spectrum users. As a result, it is inherently difficult for dynamic sharing systems to infer when and where passive receivers are vulnerable. This asymmetry strongly favors rule-based protection for passive services, and limits the effectiveness of opportunistic or sensing-only DSS schemes.

\subsection{Hybrid protection}

Given these constraints, a hybrid protection model emerges as a pragmatic and technically grounded approach. In such a framework, frequency bands that are essential to passive radio sciences remain statically allocated and strongly protected. These core bands include protected spectral lines and windows where even brief interference would irreversibly compromise scientific objectives. In these regions, dynamic access is simply incompatible with the required level of reliability.

At the same time, adjacent or extended bands may offer opportunities for carefully constrained dynamic access. In these bands, passive observations may be more flexible in frequency, time, or observing strategy. Dynamic access could be permitted under strict conditions, such as limits on power flux density, conservative geographic exclusion zones, predefined time windows, and credible enforcement mechanisms. Importantly, the existence of dynamic access in these bands must not weaken protection in the core allocations, either through out-of-band emissions or through aggregate effects.

This hybrid approach reflects the operational reality of passive sciences. Some scientific drivers are fundamentally tied to specific frequencies, while others can benefit from additional bandwidth if it can be made sufficiently quiet. Rather than treating all spectrum equally, the hybrid model acknowledges differences in scientific sensitivity and tolerance, and allocates regulatory tools accordingly.

\subsection{Comparative perspective}

Table~\ref{tab:dss_comparison} summarizes the strengths and limitations of several spectrum governance frameworks from the perspective of passive radio sciences. Static exclusive allocations offer predictability, stability, and relatively straightforward enforcement, but face increasing political pressure due to perceived inefficiency. Broad, unconstrained DSS promises flexibility and efficiency but imposes a high reliability burden and introduces multiple failure modes that are difficult to reconcile with the needs of passive services. Hybrid frameworks attempt to capture the advantages of both, preserving sanctuaries while allowing cautious expansion, but they require careful design, propagation-aware rules, and credible attribution and compliance mechanisms.

\begin{table}[h]
\centering
\begin{tabular}{p{2.7cm} p{4.3cm} p{4.3cm}}
\toprule
\textbf{Framework} & \textbf{Strengths for passive services} & \textbf{Key weaknesses and risks} \\
\midrule
Static exclusive allocation &
Predictable protection, long-term stability, simple enforcement &
Perceived inefficiency, political pressure, limited ability to exploit temporal or geographic variability \\
\midrule
Broad DSS &
Potential efficiency gains, adaptive use, applicability in dense environments &
High reliability burden, limited passive protection, multiple failure modes, and strong dependence on monitoring and trust \\
\midrule
Hybrid &
Preserves essential sanctuaries while enabling cautious expansion, adaptable to band and region &
Requires conservative propagation-aware rules, credible attribution, and increased governance complexity \\
\bottomrule
\end{tabular}
\caption{A passive-science view of spectrum governance options.}
\label{tab:dss_comparison}
\end{table}

\section{Just-in-Time Quiet Zones (JITQZ)}

Dynamic spectrum sharing frameworks often struggle to reconcile flexibility with the stringent protection requirements of passive radio sciences. In response to this tension, we propose the concept of \emph{just-in-time quiet zones} (JITQZ) as a targeted and operationally realistic mechanism for protecting high-value passive observations within otherwise dynamic spectrum environments. Rather than relying on permanent exclusions or broad static protections, JITQZs aim to provide enforceable quiet conditions precisely when and where they are most needed.

\subsection{Concept}

A just-in-time quiet zone is a temporary and enforceable reduction of the risk of RFI, activated for a specific frequency range, geographic region, and time interval. Unlike permanent radio quiet zones, which impose continuous constraints regardless of scientific activity, JITQZs are triggered on demand to support specific observations or retrieval windows with high scientific or societal value.

Operationally, a JITQZ can be described by the tuple:
\[
\mathcal{Q} = \left( \mathcal{F}, \mathcal{A}, \mathcal{T}, \mathcal{P}, \mathcal{G} \right),
\]
where $\mathcal{F}$ defines the protected frequency range or spectral mask, $\mathcal{A}$ specifies the geographic region over which constraints apply, and $\mathcal{T}$ denotes the activation time window. The term $\mathcal{P}$ captures technical emission constraints such as equivalent isotropically radiated power or power flux density limits, and $\mathcal{G}$ encodes the governance layer, including which users are subject to restrictions, how compliance is verified, and what consequences apply in the event of violations.

This formulation emphasizes that a JITQZ is not merely a technical configuration but a coupled technical and regulatory construct. Its effectiveness depends as much on governance and enforcement as on signal levels and geometry.

\subsection{Application}

The appeal of JITQZs for passive services lies in their alignment with both scientific needs and political realities. From a policy perspective, strong spectrum constraints are easier to justify when they are clearly linked to well defined public benefits and limited in duration. Temporary protections tied to specific scientific campaigns or environmental monitoring events are often more acceptable to other spectrum users than permanent exclusions, particularly when the rationale is transparent and the activation criteria are predictable.

For radio astronomy, JITQZs offer a way to protect observations that are especially vulnerable to interference. These include deep integrations that accumulate signal over many hours, time-critical campaigns targeting rare or transient phenomena, coordinated observations involving multiple observatories, and commissioning or calibration periods during which data stability is essential. In each of these cases, the scientific cost of interference is disproportionately high relative to the duration of the observation.

Similar considerations apply to passive Earth remote sensing and meteorology. Certain satellite overpasses, atmospheric conditions, or retrieval modes are particularly sensitive to interference. A JITQZ activated during these periods could protect the integrity of measurements that feed directly into weather forecasts, climate records, or disaster response systems. In this sense, JITQZs provide a mechanism for matching the level of protection to the level of societal value at stake.

\subsection{Failure modes}

Despite their conceptual appeal, JITQZs introduce their own risks. A just-in-time quiet zone is only as effective as the systems that implement and enforce it. Coordination failures can arise from delayed updates, inconsistent databases, or time synchronization errors between stakeholders. Even brief lapses can compromise observations that depend on continuous quiet conditions.

Non-compliance represents another critical failure mode. Violations may be accidental, due to misconfiguration or misunderstanding of constraints, or intentional, driven by incentives that favor transmission over compliance. Because passive observations can be disrupted by a single violator, the reliability requirements for JITQZ enforcement are exceptionally high.

Propagation effects further complicate enforcement. Anomalous propagation, such as tropospheric ducting, can extend the reach of interference well beyond planned boundaries, undermining assumptions about geographic containment. Finally, the monitoring infrastructure deployed to verify compliance may itself pose risks, particularly when placed near sensitive observatory equipment. This tension leads directly to what we refer to as the ``monitoring paradox''.

\section{``Monitoring paradox''}

Radio observatories operate under some of the strictest EMC requirements of any scientific infrastructure. On-site and nearby systems are carefully controlled through shielding, filtered power distribution, extensive use of optical fiber, meticulous grounding practices, and restrictions on consumer and industrial electronics. These measures are necessary because the signals of interest are often far below the emission levels of even well-designed electronic devices.

At the same time, many DSS proposals rely on improved spectrum monitoring to support compliance, situational awareness, and enforcement. This creates a fundamental paradox. The very instruments intended to protect passive services by monitoring spectrum use can themselves become sources of RFI if not designed and deployed with extreme care.

\subsection{Passive monitors RFI}

The label ``passive'' can be misleading when applied to spectrum monitoring equipment. Even receivers that do not intentionally transmit can generate unintended emissions \cite{Jessner2014EMI,Jackson2013EMC,ReaderEMC}. Local oscillators and their harmonics, digital clocks, switching power supplies, and high-speed data processing all produce RFI that can couple into the environment. Front-end protection stages may generate intermodulation products, and poorly shielded enclosures or cables can act as unintended antennas.

For a high-sensitivity radio telescope, these emissions do not need to be strong to be harmful. Levels that would be entirely negligible in most engineering contexts can exceed astronomical signal levels by many orders of magnitude. As a result, the deployment of monitoring equipment near observatory front ends must be approached as a full EMC engineering problem rather than as a routine instrumentation task.

\subsection{Design implications}

A monitoring architecture that is safe for passive observatories typically relies on physical separation and conservative design choices. Distance remains one of the most effective forms of mitigation, as path loss reduces the impact of unintended emissions. When monitoring is required, electronics are often housed in shielded enclosures or underground vaults, with signals transported via optical fiber to avoid conductive paths. Power supplies and clocking architectures are selected and characterized to minimize broadband emissions, and equipment is tested extensively prior to deployment.

Operational constraints are also common. Monitoring systems may be placed into reduced-activity modes during sensitive observations, or scheduled to operate only when the telescope is not observing. These measures reflect a practical reality, which is that the observatory itself must remain electromagnetically conservative at all times.

The resulting conclusion is straightforward. Enforcement of DSS and JITQZs should not depend primarily on sensors located at the telescope site. Distributed monitoring, remote sensing, and network-based compliance mechanisms can support enforcement without compromising the electromagnetic environment of observatories. In this context, protecting passive services requires not only spectrum policy innovation, but also careful systems engineering that respects the extreme sensitivity of the instruments involved.

\section{RF propagation}

DSS frameworks often rely on the assumption that interference can be predicted and geographically contained with reasonable accuracy. In practice, RF propagation is one of the most challenging and least controllable aspects of spectrum management. Terrain, atmosphere, and ionospheric conditions introduce variability that can defeat even well-intentioned protection rules.

At terrestrial scales, diffraction over terrain, scattering from clutter, and reflections from infrastructure or bodies of water can significantly extend the reach of emissions beyond nominal line-of-sight predictions. At higher altitudes, atmospheric refractivity varies with temperature, humidity, and pressure, altering propagation paths over timescales ranging from minutes to hours. One particularly important phenomenon is tropospheric ducting, in which refractivity gradients trap radio waves and allow them to propagate over hundreds of kilometers with relatively low attenuation.

At lower frequencies, ionospheric effects become dominant. Sporadic E, for example, refers to the intermittent formation of dense ionized layers in the E-region of the ionosphere. These layers can reflect or refract radio waves at frequencies well above those normally affected by ionospheric propagation, enabling unexpected propagation over long distances that can appear suddenly and persist for hours. Such events are difficult to predict and can cause interference to appear far outside any planned exclusion zone. Together, these mechanisms mean that interference is rarely confined to the neat geometric boundaries assumed in many DSS models.

\subsection{Propagation uncertainty and conservative design}

From the perspective of passive services, protection can be framed in probabilistic terms. Let $I$ denote the interference power, or equivalently the power flux density, at a passive receiver. A basic protection objective is that the probability of exceeding a harmful interference threshold remains acceptably small:
\[
\Pr\left[ I > I_{\text{thr}} \right] \le \epsilon,
\]
where $I_{\text{thr}}$ represents the maximum interference level that can be tolerated without compromising observations, and $\epsilon$ is a small probability reflecting the acceptable risk of exceedance.

In international spectrum regulation, harmful interference thresholds for passive services are not arbitrary. The ITU Radiocommunication Sector (ITU-R) has established protection criteria for radio astronomy and other passive services that explicitly relate permissible interference levels to receiver noise temperature, observing bandwidth, and integration time. In particular, ITU-R Recommendation~RA.769 defines threshold levels of interference that are considered detrimental to radio astronomical observations under a range of observing scenarios, and remains the primary reference for coordination and protection of radio astronomy services worldwide~\cite{ITU_RA769}.

Additional ITU-R Recommendations, including RA.1513 and RA.1631, further address practical aspects of protecting radio astronomy from unwanted emissions and aggregate interference, including statistical considerations and sharing scenarios~\cite{ITU_RA1513, ITU_RA1631}. Together, these documents provide a common technical baseline for coordination and regulatory decision-making.

However, in practice, individual observatories may exhibit different sensitivities, observing modes, front-end characteristics, and operational constraints. As a result, harmful interference thresholds often need to be computed on a case-by-case basis using ITU-R criteria as a reference rather than as a one-size-fits-all limit. Moreover, as instruments evolve and sensitivity improves, protection thresholds may require reassessment, underscoring the need for feedback mechanisms that allow empirical validation and adjustment when real-world interference exceeds modeled expectations.

This raises two important considerations for DSS. First, under uncertain propagation conditions, the interference $I$ should be treated as a random variable rather than a deterministic quantity. Protecting passive services therefore requires controlling not just the average interference level, but the tail of the distribution. Rare propagation events matter because a single exceedance can irreversibly damage a scientific dataset.

Second, harmful interference thresholds should not be viewed as static constants. As instruments are upgraded, observing modes change, or new mitigation techniques are introduced, these thresholds may need to be revisited. A credible DSS framework must therefore allow for feedback and revision. If real-world measurements reveal that a computed threshold is insufficiently protective, there must be a mechanism to adjust constraints rather than treating the original calculation as immutable.

Taken together, these considerations push DSS design toward conservative margins. Larger exclusion zones, stricter emission limits, and explicit safety buffers are not signs of inefficiency, but necessary responses to physical uncertainty.

\subsection{Adaptive exclusion zones}

One way to incorporate propagation uncertainty without abandoning dynamic approaches altogether is through adaptive exclusion zones. In this concept, the radius of a protected region is not fixed, but varies as a function of the prevailing propagation environment. Let $R(t)$ denote the effective exclusion radius at time $t$, expressed as
\[
R(t) = R_0 + \Delta R(\mathbf{s}(t)),
\]
where $R_0$ is a baseline exclusion radius and $\mathbf{s}(t)$ represents a vector of environmental state variables, such as meteorological conditions or refractivity profiles.

When conditions favor long-range propagation, for example during ducting events, the additional margin $\Delta R$ increases, expanding the protected region. When propagation conditions are benign, constraints can relax accordingly. Importantly, this approach does not require fragile real-time sensing at the telescope itself. Instead, it relies on conservative modeling, external environmental data, and policy rules that prioritize the protection of passive services.

The broader implication is that RF propagation is not a secondary detail that can be abstracted away in spectrum sharing discussions. It is a primary constraint that shapes what forms of DSS are feasible, how conservative protection rules must be, and where static allocations remain the only reliable option.

\section{Spectrum Awareness}

A central tension in contemporary spectrum policy is the mismatch between the ambition of dynamic spectrum sharing and the limited empirical knowledge of how spectrum is actually used. In many cases, regulators do not possess high-resolution, continuously updated maps of spectrum occupancy, particularly outside dense metropolitan areas. As a result, decisions about sharing are often made on the basis of coarse assumptions or partial datasets. In such circumstances, dynamic spectrum sharing risks becoming a solution in search of a problem, especially in rural or remote regions where the spectrum may already be naturally quiet under existing static allocations.

This lack of situational awareness is not a minor technical inconvenience. It directly affects the credibility, effectiveness, and legitimacy of DSS frameworks, particularly for services that depend on extreme sensitivity and predictable interference conditions.

For passive radio sciences, spectrum awareness is fundamentally local. The suitability of a site for a radio observatory, a remote sensing ground station, or long-term environmental monitoring depends not on national averages or nominal allocations, but on empirical spectrum measurements at that specific location, and how those conditions evolve over time.

Field measurement campaigns consistently demonstrate that spectrum occupancy and RFI levels vary dramatically with geography, local infrastructure, and propagation environment. Sites separated by tens of kilometers can experience orders-of-magnitude differences in interference, even within the same regulatory regime. Moreover, RFI environments are not static. New deployments, changes in network configuration, and evolving satellite constellations can alter local conditions on timescales far shorter than the lifetime of scientific infrastructure.

Without detailed and sustained measurements, it is difficult to distinguish between areas where spectrum sharing could meaningfully increase efficiency and areas where it would introduce unnecessary risk. For passive sciences, this distinction is critical. Once an observatory is built, its scientific productivity depends on long-term preservation of a quiet environment. Retrofitting protection after interference has become entrenched is rarely successful.

\section{Accountability via Pseudonymetry}

Dynamic spectrum sharing frameworks are inherently fragile when violations cannot be reliably attributed. When a passive service experiences harmful interference but cannot identify its source with confidence, enforcement becomes slow, uncertain, and politically contested. Investigations may span multiple jurisdictions, responsibility may be disputed among operators, and corrective actions may arrive long after the affected observations have been irreversibly compromised. In such an environment, formal protection rules alone are insufficient. What ultimately matters is whether violations can be traced to specific actors in a timely and credible manner.

This attribution problem is particularly acute for passive radio sciences. Because these services do not transmit, they cannot signal distress or negotiate access in real time. Their ability to protect themselves depends almost entirely on the external enforcement of rules imposed on transmitters. Without reliable attribution, the deterrent effect of those rules is weak, and dynamic sharing arrangements tend to erode over time as incentives favor opportunistic behavior.

Pseudonymetry \cite{Weldegebriel2025Sensing,Hellbourg2024DynamicRFI,Weldegebriel2024Watermarking,Weldegebriel2025RFI} offers a technological pathway to strengthen attribution in dynamic spectrum environments. In this context, pseudonymetry refers to the embedding of identifiable signatures directly into transmitted waveforms, allowing emissions to be associated with specific transmitters or classes of transmitters without requiring continuous disclosure of operational details. One practical approach is watermarking orthogonal frequency-division multiplexing (OFDM) signals, which are widely used in modern communication systems.

By embedding low-level, structured identifiers into the signal itself, pseudonymetry enables interference to be traced even when it is received far from the transmitter or outside nominal service areas. This capability supports several critical functions in a DSS ecosystem. First, it allows rapid identification of interfering emitters, reducing the time between a harmful event and corrective action. Second, it enables auditing of compliance with time, frequency, and power constraints by comparing observed emissions with authorized operating parameters. Third, it makes penalties and other enforcement mechanisms credible, because violations can be linked to specific actors rather than treated as anonymous or ambiguous events. Finally, it aids in diagnosing unintended emissions or misconfigurations, helping operators correct problems before they become persistent sources of interference.

From a policy perspective, the value of pseudonymetry not only lies in detection, but also in deterrence. When transmitters know that violations can be attributed with high confidence, the incentive structure changes. In game theoretic terms, pseudonymetry increases the probability that non-compliant behavior will be detected and attributed, thereby altering the payoff associated with violating sharing rules. Even modest improvements in attribution probability can shift equilibrium behavior from opportunistic violation toward sustained compliance, particularly in repeated interaction settings.

In this sense, pseudonymetry functions as a technological enabler of trust. It does not eliminate the need for clear rules, conservative protection thresholds, or robust governance. However, it provides the missing link between policy intent and enforceable reality, making dynamic spectrum sharing arrangements more credible for passive services that cannot otherwise defend themselves in contested spectral environments.

\section{Game-theoretic analysis of DSS for Passive Services}

This section develops explicit game theoretic models to clarify why informal or voluntary coexistence arrangements often fail in practice, under what conditions cooperation can be sustained, and what mechanisms are required to provide enforceable quiet for passive radio sciences in dynamic spectrum sharing environments.

\subsection{Model ingredients and interpretation}

We consider a simplified but representative setting. A single passive user, denoted $A$, represents a radio observatory or passive sensing system that derives value from low RFI access to the spectrum. A set of $N$ active users, indexed by $i \in \{1,\dots,N\}$, represents transmitting systems such as commercial networks, satellite operators, or federal systems. Time is divided into discrete intervals $t = 1,2,\dots$, during which a just-in-time quiet zone (JITQZ) may be activated.

Several parameters characterize the incentives faced by each actor. The quantity $V_A$ denotes the value to the passive service of a successfully protected interval. This value may reflect the scientific return of a unique observation, the avoided cost of repeating an observation, or the preservation of data integrity. Conversely, $L_A$ represents the loss incurred by the passive service if harmful interference occurs during a protected interval. In many cases, this loss is effectively irreversible.

For each active user $i$, $G_i$ denotes the gain from transmitting during a quiet interval. This gain may correspond to additional revenue, improved quality of service, mission completion, or operational convenience. The cost of compliance, $C_i$, captures the burden imposed by respecting quiet-zone constraints, such as reducing transmit power, shifting to a different band, delaying transmissions, or reconfiguring a network.

Two regulatory parameters govern enforcement. The probability $p$ represents the likelihood that a violation is detected and correctly attributed to a specific actor. The penalty $F_i$ represents the consequence of being caught violating, which may take the form of a fine, loss of access privileges, reputational damage, or regulatory sanctions. Together, $p$ and $F_i$ determine whether compliance is a rational choice for active users.

We model the decision of each active user during a quiet interval as a binary action $a_i \in \{0,1\}$, where $a_i = 1$ denotes compliance with the quiet-zone constraints and $a_i = 0$ denotes violation.

\medskip
\noindent\textbf{Summary of model quantities:}
\begin{itemize}[leftmargin=*]
    \item $A$: passive user (e.g., radio observatory or passive sensing system).
    \item $i \in \{1,\dots,N\}$: index over active users (transmitting systems).
    \item $t$: discrete time interval during which a JITQZ may be active.
    \item $V_A$: value to the passive service of a successfully protected interval.
    \item $L_A$: loss to the passive service if harmful interference occurs.
    \item $G_i$: gain to active user $i$ from transmitting during a quiet interval.
    \item $C_i$: cost to active user $i$ of complying with quiet-zone constraints.
    \item $p$: probability that a violation is detected and correctly attributed.
    \item $F_i$: penalty incurred by active user $i$ if a violation is detected.
    \item $a_i \in \{0,1\}$: action of active user $i$ (1 = comply, 0 = violate).
\end{itemize}

\subsection{One-shot externality game}

Consider first a one-shot interaction in which a JITQZ is announced for a single interval. Each active user compares the payoff from complying with the payoff from violating. Compliance yields a payoff of $-C_i$, reflecting the cost of foregone transmission. Violation yields a payoff of $G_i - pF_i$, reflecting the gain from transmission offset by the expected penalty.

A rational active user will choose to violate whenever
\[
G_i - pF_i > -C_i,
\]
or equivalently,
\[
G_i + C_i > pF_i.
\]

If the probability of detection is low, or if penalties are weak, violation dominates compliance for many actors. Crucially, from the perspective of the passive service, this is a weakest-link problem. Even if most active users comply, a single violator can ruin the protected interval. As a result, voluntary coexistence is inherently fragile, and protection fails unless the system induces universal compliance.

\subsection{Repeated interaction and future access}

In reality, spectrum access decisions are repeated over time. Active users typically care about future access opportunities, not just immediate gains. This can be modeled by introducing a discount factor $\delta \in (0,1)$, which captures how much an actor values future payoffs relative to immediate ones.

Suppose that violations trigger a punishment phase, such as loss of dynamic access privileges for $T$ future intervals, or the imposition of stricter constraints. Let $P_i$ denote the expected discounted loss associated with this punishment. $P_i$ depends on $\delta$, and actors who value long-term access highly (high $\delta$) experience larger effective penalties.

In a repeated setting, violation becomes unattractive when
\[
G_i + C_i < pF_i + pP_i.
\]
Compared to the one-shot case, the term $pP_i$ represents the expected future cost of being caught. This term can dominate immediate incentives if punishment is credible and future access is valuable.

The implication for passive protection is clear. Repeated game cooperation can only be sustained when detection and attribution are reliable, when punishments are meaningful, and when active users care sufficiently about future access. Without these conditions, repeated interaction alone does not solve the interference problem.

\subsection{Quiet-zone bargaining}

Dynamic spectrum sharing is often described as a source of win--win outcomes. This claim can be formalized by considering the aggregate welfare change associated with granting a quiet interval of duration $\tau$:
\[
\Delta W(\tau) = V_A(\tau) - \sum_{i=1}^N C_i(\tau).
\]

A necessary condition for a Pareto-improving outcome is that the value of the quiet interval to the passive service exceeds the total compliance cost borne by active users. Even when this condition holds, individual active users may still lose unless costs are redistributed. This creates a role for side payments, spectrum credits, or other compensation mechanisms.

The value $V_A$ is however rarely monetized in a way that is comparable to commercial gains. Scientific value is often long term, societal, and non market in nature. As a result, purely market-based bargaining mechanisms may systematically undervalue passive services unless supported by explicit policy intervention.

\subsection{Incentives, leadership, and attribution}

One approach to improving compliance is a credit-based system, in which active users earn credits for respecting quiet zones, and lose them when violating. Credits can translate into future access priority, wider bandwidth, or relaxed constraints during non-sensitive periods. Let $K_i$ denote the expected future value of such credits, and let $\Delta K_i$ denote the credit loss associated with a detected violation. Compliance is then favored when
\[
G_i + C_i < K_i + pF_i + p\Delta K_i.
\]
Credit systems are particularly attractive when direct fines are politically difficult or administratively costly.

More generally, DSS can be viewed as a Stackelberg game in which the regulator acts as a leader, selecting enforcement parameters such as monitoring investment (which affects $p$), penalties $F_i$, and quiet-zone rules, while active users respond by choosing whether to comply. In principle, reliable protection requires that the regulator be able to ensure
\[
pF_i \ge G_i + C_i
\]
for the relevant classes of actors. When this condition cannot be met, passive protection is not robust, regardless of how well-intentioned the sharing framework may be.

This is where attribution technologies play a decisive role. Increasing the probability that violations are detected and correctly attributed is often more effective than escalating penalties. Pseudonymetry directly targets this lever by increasing $p$ and reducing enforcement latency. A simple representation is
\[
p = p_0 + \Delta p_{\text{pseudo}},
\]
where $p_0$ is the baseline detection probability and $\Delta p_{\text{pseudo}}$ reflects the improvement enabled by watermarking and identification infrastructure. Even modest increases in $p$ can shift equilibrium behavior when the system operates near the compliance threshold.

\subsection{Safety-critical protection}

Unlike many commercial sharing scenarios, passive radio sciences do not degrade gracefully under interference. If a quiet interval is compromised when any one of $N$ actors violates, and violations occur independently with probabilities $q_i$, the probability of success is
\[
\Pr[\text{quiet success}] = \prod_{i=1}^N (1 - q_i).
\]
As the number of potential interferers grows, maintaining a high probability of success requires each $q_i$ to be extremely small. This scaling behavior explains why DSS for passive services resembles safety engineering rather than conventional spectrum optimization. Strong constraints on participation, conservative margins, and robust enforcement are not optional.

\subsection{Information asymmetry and limits of regulation}

All of the mechanisms discussed above assume some degree of shared knowledge about who is transmitting, where, at what power, and under what propagation conditions. In practice, regulators and operators often lack this state information. This imbalance creates two problems. Some actors may be tempted to ignore the rules when violations are unlikely to be detected, while others who are already willing to behave responsibly are the ones most likely to participate in sharing frameworks.

The result is fragile equilibria that can collapse when trust erodes. Improving spectrum awareness through independent monitoring, mobile sensing, and transparent data sharing directly addresses this limitation by increasing the common knowledge available to all parties. Without such investment, even well-designed DSS frameworks may fail to deliver reliable protection for passive services.

\section{Engineering and governance requirements for passive-safe DSS}

The preceding sections make clear that protecting passive radio sciences in a dynamic spectrum sharing environment is not a matter of goodwill or high level policy intent. It is an engineering and governance problem with explicit reliability requirements. This section translates the analysis into a set of concrete criteria against which any DSS framework claiming to protect passive services should be evaluated. These requirements are cumulative rather than substitutable, as failure in any one dimension undermines the entire protection framework.

\subsection{Reliability requirements}

At a minimum, a passive-safe DSS framework must articulate explicit reliability targets. Protection cannot be expressed solely in terms of average interference levels or nominal compliance with emission limits. Instead, it must specify acceptable tail risk, for example by bounding the probability that interference exceeds a harmful threshold, $\Pr[I > I_{\text{thr}}]$. This reflects the reality that rare events, not averages, dominate the risk for passive observations.

In addition, DSS systems must define fail-safe behavior. Databases may become unavailable, sensing inputs may be corrupted, and coordination links may fail. In such cases, the system should default to conservative behavior that favors protection rather than access. Time synchronization and update latency must also be explicitly budgeted. For observations that require continuous quiet conditions, even brief mismatches between authorization state and actual transmissions can be damaging.

\subsection{Propagation-aware margins}

Any credible DSS framework must explicitly account for uncertainty in RF propagation. Protection rules based on nominal or median propagation conditions are insufficient when rare atmospheric or ionospheric events can extend interference over long distances. Conservative assumptions are therefore not optional, they are a direct consequence of physical uncertainty.

Where dynamic approaches are used, exclusion zones and emission limits should be adaptive, expanding under conditions that favor long-range propagation, and relaxing only when supported by robust evidence. Crucially, these models must be validated against real world measurements. Field data provide the only reliable check on whether assumed margins are adequate, and they should inform periodic reassessment of protection rules.

\subsection{Monitoring and attribution}

Monitoring is necessary for enforcement, but it must be implemented in a way that does not compromise the EMC of observatory sites. Protection schemes that rely on placing active or poorly characterized equipment near sensitive receivers are fundamentally flawed.

Beyond monitoring, attribution is essential. Compliance cannot be enforced if violations cannot be linked to specific actors. DSS frameworks should therefore include auditable logs, transparent compliance metrics, and attribution mechanisms that are technically credible. Where appropriate, waveform-level techniques, such as pseudonymetry, can provide a direct link between observed interference and responsible transmitters, strengthening both enforcement and deterrence.

\subsection{Incentives and enforcement}

Finally, protection must involve incentive. The enforcement regime must ensure that compliance is a rational choice for active users. In the language developed earlier, penalties and the loss of future privileges must be sufficient to satisfy
\[
G_i + C_i < pF_i + pP_i
\]
for the relevant classes of actors. This requires not only formal penalties, but also credible detection and timely attribution.

Enforcement mechanisms should be as automatic and predictable as possible. Discretionary or ad hoc enforcement undermines deterrence and erodes trust over time. At the governance level, protection commitments must be robust across political cycles. Passive scientific infrastructure often operates on decades timescales, and its protection cannot depend on short-term policy priorities.

\section{DSS vs. static allocations}

The analysis above does not lead to a binary conclusion in favor of or against DSS. Instead, it points to a conditional landscape in which DSS may be appropriate in some contexts, and fundamentally unsuitable in others. Importantly, the criteria below should be viewed as jointly necessary. Satisfying one or two conditions is not sufficient to justify dynamic access.

\subsection{DSS as an opportunity}

DSS may offer value for passive sciences when several conditions are simultaneously met. The frequency band in question should not correspond to a critical protected spectral line, and passive observations in that band should exhibit some flexibility in frequency or timing. The set of active users should be limited, identifiable, and subject to effective governance.

Equally important, enforcement and attribution mechanisms must be credible, resulting in a high probability of detecting and attributing violations. Finally, propagation risk must be manageable through conservative margins that have been validated empirically. Only when all of these conditions are satisfied does DSS have a realistic chance of providing net benefit without unacceptable risk.

\subsection{DSS as a risk}

Conversely, static allocations and strong, permanent protections remain indispensable when observations require extremely low interference over long integration times, when the number of potential interferers is large, or when propagation conditions can produce rare but severe interference events. In such cases, the weakest-link nature of passive protection dominates, and even a single failure can negate an entire observation.

Static protection is also essential when the policy environment cannot sustain strong enforcement over time. Where attribution is weak, penalties are uncertain, or governance commitments are unstable, dynamic sharing frameworks tend to degrade in practice, leaving passive services exposed.

\subsection{Tiered protection framework}

To reconcile competing pressures, we propose a tiered protection framework that aligns regulatory tools with scientific sensitivity and risk tolerance.

At the foundation, Tier~0 consists of permanent passive sanctuaries. These include core protected bands, key spectral lines, and established radio quiet zones. In these regions, static allocations and strong exclusions remain non-negotiable.

Tier~1 encompasses conditional dynamic coexistence in adjacent or extended bands. In this tier, dynamic access may be permitted under tightly controlled conditions, including just-in-time quiet zones, conservative power limits, adaptive exclusion regions, and robust enforcement. The defining feature of this tier is that dynamic access is subordinate to passive protection and can be withdrawn when conditions are not met.

Tier~2 includes opportunistic access regimes primarily oriented toward commercial efficiency, where passive constraints are minimal or absent. These regimes are appropriate only where passive use is negligible or non-existent and should not be allowed to encroach on Tier~0 or Tier~1 protections through aggregate effects or regulatory drift.

This tiered approach provides a structured way to integrate DSS into existing regulatory frameworks without eroding the protections that passive radio sciences require. It acknowledges that not all spectrum is equal, that not all risks are acceptable, and that different bands demand different governance tools.

\section{Conclusion}

Dynamic spectrum sharing is an increasingly important direction in spectrum policy, and in some contexts it can deliver genuine gains in efficiency and flexibility. For passive radio sciences, however, DSS is neither a universal solution nor an automatic improvement. It introduces new classes of failure modes, depends critically on accurate modeling of radio frequency propagation, and shifts the burden of protection toward real time coordination, monitoring, and enforcement. These requirements are substantially more stringent than those faced by most active services.

The concept of just-in-time quiet zones provides a pragmatic bridge between static protection and dynamic access. By tying strong, enforceable constraints to well-defined observation windows with clear scientific or societal value, JITQZs align better with both the operational needs of passive services and the political realities of spectrum governance. Within a hybrid framework, where core passive allocations and radio quiet zones remain permanently protected, and dynamic access is considered only in adjacent or extended bands, JITQZs offer a path toward limited coexistence without eroding essential safeguards.

A central result of the game-theoretic analysis is that passive-safe DSS cannot rely on voluntary compliance or informal coordination. Protection is only reliable when compliance constitutes a rational equilibrium for active users. Achieving this requires a sufficiently high probability of detection and attribution, coupled with consequences that are meaningful on operational timescales. Attribution technologies such as OFDM watermarking for pseudonymetry play a critical enabling role by increasing enforcement credibility and reducing latency, thereby shifting incentives away from opportunistic behavior and toward sustained cooperation.

At the same time, the analysis highlights a more fundamental limitation. Dynamic sharing frameworks presuppose a level of spectrum situational awareness that often does not exist today, particularly outside dense urban environments. Without sustained investment in spectrum measurement infrastructure, including distributed and mobile sensing approaches, DSS risks being deployed without a reliable empirical foundation. In such cases, sharing may address perceived inefficiencies while inadvertently increasing risk for services that depend on exceptionally quiet spectral environments.

The overarching conclusion is therefore conditional. Dynamic spectrum sharing can be an opportunity for passive radio sciences only when it is designed as a high-reliability, enforceable quiet system. Static allocations remain indispensable for the most sensitive observations and for long-term scientific continuity. Dynamic techniques should be deployed selectively, guided by empirical measurements, conservative propagation-aware safeguards, rigorous EMC practices, and governance structures capable of sustaining accountability over decades. Anything less risks trading short-term flexibility for long-term scientific loss.


\end{document}